# Dynamics of Viscous Entrapped Saturated Zones in Partially Wetted Porous Media

Shuoqi Li [1], Mingchao Liu [1,2], Dorian Hanaor [3] and Yixiang Gan [1,*]

[1] School of Civil Engineering, The University of Sydney, Sydney, NSW 2006, Australia
[2] Department of Engineering Mechanics, CNMM & AML, Tsinghua University, Beijing 100084, China
[3] Chair of Advanced Ceramic Materials, Technische Universität Berlin, Germany

## Abstract

As a typical multiphase fluid flow process, drainage in porous media is of fundamental interest in nature and industrial applications. During drainage processes in unsaturated soils and porous media in general, saturated clusters, in which a liquid phase fully occupies the pore space between solid grains, affect the relative permeability and effective stress of the system. In this study, we experimentally studied drainage processes in unsaturated granular media as a model porous system. The distribution of saturated clusters is analysed by an optical imaging method under different drainage conditions, in which pore-scale information from Voronoi and Delaunay tessellation was used to characterise the topology of saturated cluster distributions. By employing statistical analyses, the observed spatial and temporal information of multiphase flow and fluid entrapment in porous media are described. The results indicate that the distributions of both the crystallised cell size and pore size are positively correlated to the spatial and temporal distribution of saturated cluster sizes. The saturated cluster size is found to follow a lognormal distribution, in which the generalised Bond number ($Bo^*$) correlates negatively to the scale parameter ($\mu$) and positively to the shape parameter ($\sigma$). These findings can be used to connect pore-scale behaviour with overall hydro-mechanical characteristics in partially saturated porous media, using both the degree of saturation and generalised Bond number.

# 1. Introduction

Multiphase fluid flow in porous media is an important phenomenon with a broad range of applications, including risk assessment of landslides, biophysical systems, and groundwater hydrology (Dai et al., 2002; Kueper and Frind, 1988; Saba and Illangasekare, 2000; Seymour et al., 2004; Simmons et al., 1999). Predicting the properties of such systems usually requires pore-scale information including the topological features and interactions between intrinsic phases. In a multiphase flow system, as the wetting phase is withdrawn and replaced by the non-wetting phase (gas), the interface of wetting and non-wetting fluid phases tend to recede to the outlet as a continues moving front. The displacement dynamics of the interface masks the complex interplay of viscous, capillary, and gravitational forces in the system and attracted extensive studies (Aker et al., 2000; Bear and Braester, 1972; Bryant and Blunt, 1992; Lenormand and Zarcone, 1985; Måløy et al., 1985; Savani et al., 2017; Schwarze et al., 2013; Youngs, 1960).

Models used to study drainage processes in porous media include capillary tubes (Dullien et al., 1986; Morrow, 1970; Youngs and Aggelides, 1976) and two dimensional Hele-Shaw cells (Haines, 2009; Méheust et al., 2002). At high flow rates or in the absence of gravity, viscous forces tend to destabilise the continuous front into narrow fingers against the stabilising effect of gravitational force, which are the prerequisite conditions for the generation of limited saturated zones behind the continuous front (Dullien et al., 1986; Måløy et al., 1985; Méheust et al., 2002; Morrow, 1970; Youngs, 1960; Youngs and Aggelides, 1976). However viscous entrapped saturated zones left behind the moving front were neglected in these studies. As such, understanding interface dynamics between different liquids in drainage processes requires a better description of multiphase flow in porous media.

Fluid-fluid interface displacement in porous media along the front can be described at the pore scale as individual jumps at pore throats (Haines, 2009; Lenormand and Zarcone, 1989). Based on this, different approaches have been taken to explain fluid front morphologies, including a coupled capillaries model (Hoogland et al., 2015), which is established by a pair of neighbouring capillaries with different radii; a model consisting of small drops sliding down from different substrates (Podgorski et al., 2001); and the use of sliding liquid elements (Su et al., 2004). The competition between viscous, capillary, and gravitational forces during drainage in a two-dimensional synthetic porous medium was considered by Auradou et al. (1999) and Meheust et al. (2002) to deduce interface movements. For cases of two or more immiscible fluid phases in porous media, variations of displacement structure are governed by the physical properties of the involved fluids and porous media (Meakin et al., 1986; Paterson, 1984). These properties include the flow rates, wetting properties, viscosity ratio and density of the involved fluids, and pore structure (Auradou et al., 1999; Podgorski et al., 2001; Su et al., 2004).

The formation of drainage fronts (i.e., the interfaces between wetting and non-wetting fluids) and displacement dynamics, shaped by the aforementioned properties, determine the residual liquid phase distribution in partially wetted porous media and affect the relative permeability of the correlated unsaturated regions (Auradou et al., 1999; Scherer et al., 2007). Recent experimental and numerical studies concerning drainage processes in porous media tend to focus mainly on the evolution of viscous fingering at various flow rates and flow velocity distributions along the continuous moving front (Alim et al., 2017; Holzner et al., 2015; Lenormand and Zarcone, 1989; Matyka et al., 2016; Vishnudas and Chaudhuri, 2017; Zhao et al., 2016). However, studies on the morphological characteristics of liquid phase entrapment behind the front remain relatively scarce. Accurate predictions of the fraction of liquid trapped behind the drainage front and its distribution are essential in order to quantify the relative permeability and the effective stress of unsaturated soils and rocks (Blunt et al., 2002; Simmons et al., 1999).

For low flow rate conditions, fluid invasion events along the displacement front occur pore by pore, and can be predicted by the classic percolation theory (Gouyet et al., 1988; Hunt et al., 2014). However, for high flow rate cases, pores are invaded simultaneously, exhibiting large pressure bursts with liquid ejection. This phenomenon is manifested primarily through the complex interplay of three forces within the system, namely gravitational force, viscous force and capillary force (Or, 2008). To quantify the contribution of these three forces to viscous unstable flows in porous media, two typical dimensionless parameters, i.e., Capillary number ($Ca = \mu v a^2 / \gamma k$) and Bond number ($Bo = \Delta \rho g' a^2 / \gamma$), are defined (Al-Fossail and Handy, 1990; Alvarez et al., 2009; Birovljev et al., 1991; Blom and Hagoort; Gawlinski and Stanley, 1981; Méheust et al., 2002; Or, 2008), where $\mu$ is the viscosity of the wetting fluid, $v$ the mean front velocity, $\gamma$ the surface tension, and $k$ the permeability of the porous medium, $\Delta \rho$ the density difference between the wetting and non-wetting fluids, $g'$ the component of gravitational acceleration and $a$ is the typical pore size (more details can be found in Crisp and Williams, 1971). It is clear that the Capillary number describes the ratio of viscous to capillary forces, while the Bond number indicates the ratio of gravitational to capillary forces. Following Meheust et al. (2002), these two dimensionless parameters can be combined as the generalised Bond number, $Bo^* = Bo - Ca$ (Méheust et al., 2002; Moebius and Or, 2014), to quantify the transition between different flow regimes, i.e., stable and unstable flow. Specifically, positive values of $Bo^*$ indicate stable moving fronts while negative values of $Bo^*$ mark unstable liquid fronts.

The flow-rate dependency of the displacement front morphology raises the following question: how does the flow-rate affect the liquid phase fraction trapped behind a moving drainage front? Wildenschild et al. (2005) used X-ray tomography to image air invasion into the saturated sandy soil and observed that large values of residual saturations occurred for high flow rates. The same trend was observed in glass bead micro-models in Løvoll's study (2011). The observed increase in residual saturation behind the displacement front for higher flow rates occurs due to the preferential drainage of large pores bypassing and entrapping liquid phase in non-invaded volumes (Hoogland et al., 2015). It should be noted that liquid entrapment behind a rapidly moving front is not permanent, rather this liquid drains through a redistribution flow, albeit at a rate significantly lower than that of the primary moving front. This redistribution and the dominant liquid flow both affect the formation and evolution of these discontinuous entrapped liquid phase, which are termed locally saturated clusters. However, the extent and dynamics of locally saturated clusters are ripe for further exploration. An improved understanding of the behaviour drained and trapped liquid fractions in multiphase flow in porous media is of great value towards diverse engineering applications.

This study aims to establish a link between the complex mechanisms inside porous media and the morphological characteristics of a discontinuously entrapped liquid phase (i.e., saturated clusters). The properties of "viscous delayed" regions are studied in this work by establishing an experimental system where saturated clusters are obtained by the forced drainage of water from a porous medium. In this system, the gravitational and viscous forces can be tuned by changing tilting angles and drainage rates, and the competition of capillary, gravity and viscous forces can be examined by introducing the generalised Bond number ($Bo^*$) to mark the onset of unstable flows. In the present paper we describe an experimental setup for the acquisition of entrapped liquid profiles and present data analysis methods to determine the influence of pore-scale structure on the spatial extent of saturated clusters and their size distributions under various drainage scenarios. Results are interpreted with a view towards their significance in partially saturated porous systems.

## 2. Methodology

### 2.1 Experimental setup

The present experimental system was designed for temporally and spatially resolved imaging of interface motion in micro-models consisting of glass beads. The key elements of the setup are shown in Fig. 1(a). A micromodel was constructed by a single layer of glass beads with a diameter of 2 mm that were randomly arranged between two parallel glass plates. The plates and beads were then slightly compressed by clamps, and the model was sealed with silicone glue along the bottom and side edges, leaving the top side open to the atmosphere. The observable area was 180 mm×270 mm, and the resulting porosity was about $n \approx 0.4$. After the model was installed, the outlet channel was connected to a high precision pump (Masterflex 77200-62, USA) to provide a stable volumetric injection and withdrawal flow rate. A digital camera (Nikon D3300) controlled by a computer over digiCamControl software was used to record the experiments at given time intervals based on different flow rates. Each image was rescaled to 1000×1500 pixels, which corresponds to a spatial resolution of 30 pixels 1 mm$^2$ (~ the pore area). An electroluminescent panel provided sufficient background illumination for different shutter speeds. Both the light source and model could be tilted simultaneously to various inclination angles to control the gravitational component of the fluid flow in the porous medium. The component of gravity along the model $g'$ is given by $g \sin(\alpha)$, where $g$ is the gravitational acceleration and $\alpha$ is the inclination angle.

Water was used as the wetting fluid in all experiments, dyed with tracer (Cole-Parmer 00298-06, 2.6ml/100ml), and air was the non-wetting fluid invading from the open top edge. Experimental values of the wetting fluid were measured at 25°C. The surface tension between two fluids $\gamma$ is 64.27 mN/m (standard deviation, std=0.15 mN/m), which was performed by a goniometer (Rame-hart, model 200 Standard Contact Angle Goniometer with DROPimage Standard) with the pendent drop method. The wetting fluid has a viscosity $\mu$ of 0.95 mPa.s (std=0.2 mPa.s) and a density $\rho$ of 998 kg/m3. The permeability of the medium $k$ was estimated from the Kozeny-Carman relation as 4.89×10$^{-9}$ m$^2$ (Scheidegger, 1958). The contact angle for the container wall is 44.56° (std=0.18°), which has similar wettability to glass plates. The applied flow rates ranged from 1 ml/min to 20 ml/min for different tilt angles. These resulted in the mean front velocity determined from volumetric flow rates at the cross-section area taking into account the porosity, as shown in Table 1.

**Table 1.** Evaluation of mean front velocity and generalized bond number ($Bo^*$) used for experiments.

| Tilt Angle (°) | 0 | | | | | 5 | | | 10 | | 15 |
|---|---|---|---|---|---|---|---|---|---|---|---|
| Drainage Rate (mL/min) | 1 | 2 | 5 | 10 | 20 | 1 | 5 | 20 | 5 | 20 | 10 |
| Mean Front Velocity (mm/s) | 0.116 | 0.231 | 0.579 | 1.157 | 2.315 | 0.116 | 0.579 | 2.315 | 0.579 | 2.315 | 1.157 |
| $Bo^*$ (-) | -1.556 ×10⁻⁴ | -3.111 ×10⁻⁴ | -7.778 ×10⁻⁴ | -0.002 | -0.003 | 0.006 | 0.005 | 0.003 | 0.010 | 0.009 | 0.016 |

## 2.2 Image processing

The grayscale pixel-intensity histogram of raw images shows two distinguishable peaks corresponding to water and air phases. In order to separate the air-filled region from the saturated region, all grey levels under a given threshold were set to black ('wet') while the others were white ('dry'). After thresholding, the whole observation area was separated into different regions with glass beads assigned to wet or dry regions. Here only the saturated regions that are larger than the projection area of a typical glass bead ($\pi$ mm² in this study) were considered. After removal of trapped liquid volumes with projection areas smaller than a single glass bead, the resulting black and white images provide the required geometrical information. To allow the visualisation of separate clusters of different sizes, analyses of morphological components have been carried out and colours were assigned to the localised saturated clusters, as shown Fig. 1(b) and Fig. 1(c). The inner rings in solid beads are shown in Fig. 1(d). Small amounts of water are trapped in the void spaces between glass beads and the front and back walls. Based on the micro-model geometry, the pore space in our sample can be illustrated in two ways: a 3-D sphere network and a 2-D cylinder network. These entrapped liquid rings are discarded in this study.

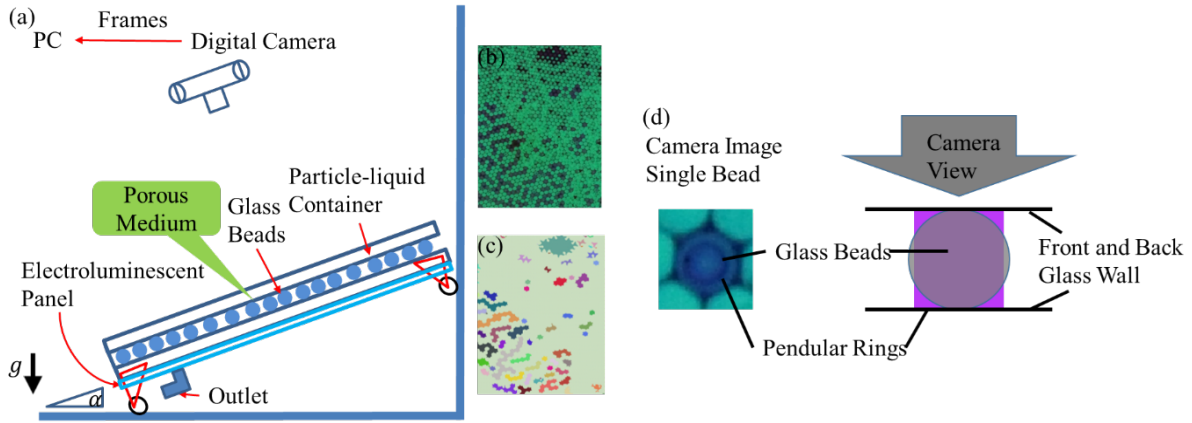

**Fig. 1** Schematic illustration of the experimental setup. (a) The camera, electroluminescent panel and cell are all connected to a common frame that can be tilted. The liquid outlet on the bottom is connected to a syringe pump that injects and withdraws the liquid. (b) A detailed zoomed-in section from the raw image extracted from the camera showing the structure of the porous network, the non-wetting phase (green) and wetting phase (dark red). (c) The same section after applying image processing. Individual saturated clusters are coloured randomly to allow the visualisation of separated clusters for different sizes. (d) Pendular rings between the glass bead and the front glass and schematic of a glass bead with pendular rings.

To consider the influence of pore structures on saturated clusters, Voronoi and Delaunay meshes of the grain packing are generated from an individual image under saturated conditions, shown in Fig. 2. Fig. 2(a) illustrates the algorithms for creating the two mesh types: red solid links show the Delaunay triangulation which is constructed on the basis of the distances between neighbouring beads, while blue dashed links are the Voronoi lattice which identifies the set of points in the space which are closest to the centre of each cell. The top inset small images in Fig. 2(b) are zoomed-in cropped images from the Voronoi lattice. The colour map is used to display the values of void ratio ($e$) of individual cells. In Fig. 2(b), $e_{max}$ has the value of 7.169 and $e_{min}$ is 0.103 as the theoretical limit for the hexagonal packing. The upper images in Fig. 2(b) show two typical crystallised and non-crystallised regions as parts of the larger lower image showing the overall experimental cell. This packing effect influences the resulting saturated cluster distribution. With the same aforementioned image processing methods, the centroids of glass beads are extracted, which are the vertexes of each triangle. Similar to the previous analysis for Fig. 2(b), the largest void ratio for the Delaunay mesh is $e'_{max} = 11.162$; while the smallest void ratio is $e'_{min} = 0.103$, shown in Fig. 2(c). In Fig. 2(c), a pore is represented by the region inside each Delaunay triangle. To yield the void area, $\pi/2$ mm$^2$ is subtracted from the area of each triangle. In this manner, the Delaunay triangulation data can be transformed into a pore size distribution curve. Similarly, the Voronoi polygon area data stream can be transformed into cell-void size distribution curve.

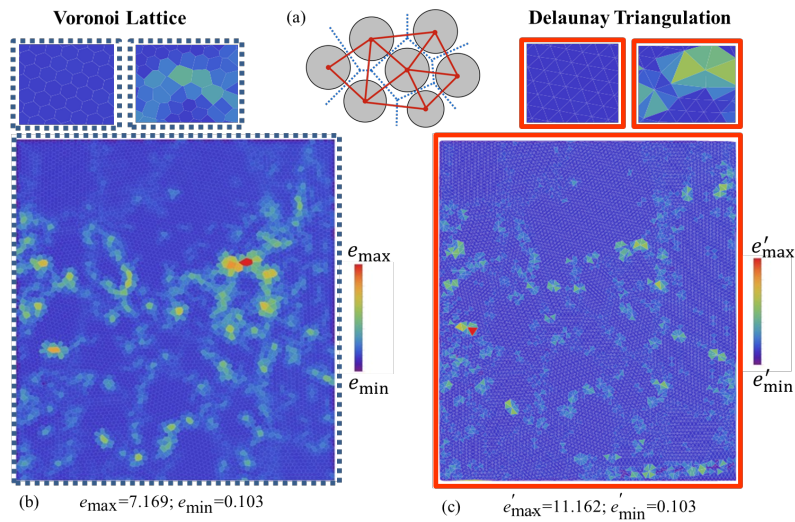

**Fig. 2** (a) Schematics of Voronoi lattice (blue dashed lines) and Delaunay triangulation (red solid lines). (b) Map of Voronoi mesh for an individual fully saturated image. The bar legend is scaled in different gradient, ranging from smallest ($e_{min}$ = 0.103) to largest ($e_{max}$ = 7.169) Voronoi cell area. (c) Map of Delaunay mesh for an individual fully saturated image. The bar legend is scaled in different gradient, ranging from smallest ($e'_{min}$ = 0.103) to largest ($e'_{max}$ = 11.162) cell void ratio.

## 2.3 Statistical analysis

In this section, we use the probability density function (PDF), $f(y|w)$, specifying the data vector $y$ under the given parameter $w$, which defines the shape of the distribution profile. Here the subscript letter for the vector element $y$ denotes statistically independent observations. According to the theory of probability, the PDF for $y = (y_1 \ldots y_m)$ can be expressed as a multiplication of their PDFs. The probability density function of the lognormal distribution considering two parameters, which describes the relative likelihood for the continuous random variable to take on a given value, is given as (Aitchison and Brown, 1976)

$$f(X|\mu, \sigma^2) = \frac{1}{\sqrt{(2\pi\sigma^2)}X} exp\left[-\frac{(\ln X - \mu)^2}{2\sigma^2}\right], \tag{1}$$

where the scale parameter $\mu$ is interpreted as the logarithmic mean of the random variables, and the shape parameter $\sigma$ is interpreted as the standard deviation of the random variables in a logarithmic scale. As the parameters change in value, different probability distributions are generated. In this case, parameter estimation, a procedure of finding parameter values of a model that best fits the data, is vital to the following analysis.

In order to find the statistical parameters that best fit the acquired experimental dataset, the maximum likelihood estimation (MLE) method stands out for its attributes of sufficiency, consistency and efficiency (Myung, 2003). By taking the product of the probability densities of the individual $X_i$ ($i = 1, 2, 3, \ldots n$), we derived the likelihood function of the lognormal distribution for a series of $X_i$. The likelihood function $L(\mu, \sigma^2|X)$ is defined by reversing the roles of data vector and parameter vector: $L(\mu, \sigma^2|X) = \prod_{i=1}^{n}[f(X_i|\mu, \sigma^2)]$. By taking the natural logarithm of the likelihood function and taking the gradient of $L$ with respect to $\mu$ and $\sigma^2$ while setting it equal to zero, we can find $\mu$ and $\sigma^2$ which maximise $L$ (Ginos, 2009). In addition, to measure the linear relationship between variables, the Pearson's correlation coefficient $r$ is employed. Moreover, the correlation coefficient is a rescaled covariance: $r_{X,Y} = \frac{cov(X,Y)}{\sigma_X \sigma_Y}$, where $cov(X, Y)$ is the sample covariance, and $\sigma_X$ and $\sigma_Y$ are sample standard deviations. After rescaling the covariance, it ranges between -1 and +1, where 1 is total positive linear correlation, 0 is no linear correlation, and -1 is total negative linear correlation (Lee Rodgers and Nicewander, 1988).

## 3. Results analysis

### 3.1 Characterisation of liquid occupied regimes

The flow of two immiscible fluids through a granular medium depends not only on the complex interplay between viscous, capillary and gravitational forces, but also the topological characteristics of the porous structure, here in particular the packing characteristics of the grains. The interaction between these forces and the pore structure yields a complex flow regime, shown in Fig. 3. It can be observed that there are four different liquid content forms, namely adsorbed, pendular, funicular and fully saturated regimes. The projection areas of these four regimes can be separated on the basis of their water content. In these four regimes, the water fraction increases from the top zoomed-in picture to the bottom one. Specifically, in the absorbed regime, liquid phase exists as thin films coating on each spheres. As the liquid phase increase, pendular bridges form between consecutive spheres. When further water is added, the liquid phase becomes continuous, while the air phase becomes discontinuous, marking the funicular regime. Finally, in the fully saturated regime, all voids are occupied by the liquid phase. In the following sections, we focus on the spatial and temporal evolution of the fully saturated regime. Though the information from other regimes may be practically useful, e.g., the percolation of the liquid bridges presented in the pore space, their study may require higher resolution at the local scale.

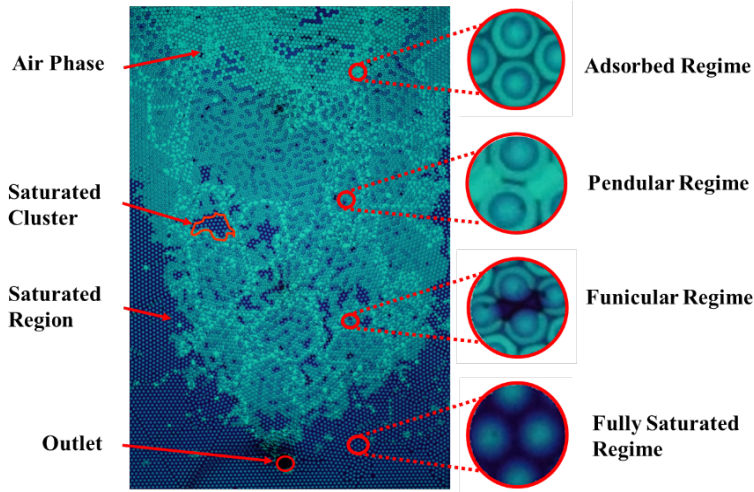

**Fig. 3** An example experimental image revealing different saturation regimes

## 3.2 Spatial distribution of wetted cells

To look into the influence of packing characteristics on saturated cluster formation and evolution during drainage processes, the correlation between the temporal-spatial information of saturated clusters and packing characteristics is compared here. Fig. 4 shows the spatial distributions of wetted cell density in two directions of the model for three flow conditions at two saturation levels. To remove boundary effects, the top edges of the model (about 20 grain diameters) are eliminated in our study. These spatial distributions are produced via a histogram of the data with linear binning of fully saturated Voronoi cells. All the points shown in Fig. 4 are the histogram heights with unit bin width. The wetted cells of interest in our analysis are those located inside saturated clusters.

In Figs. 4(a) and 4(b), we divided the top-to-bottom distance into 45 equal sub windows (with heights of 5 mm), each one spanning the whole porous medium width (230 mm). The histograms show the probability density of wetted cells within each sub window along the $y$-coordinates, the distance between the sub window centres and the outlet. These measurements are taken at the time points when the liquid occupies 50% and 60% of the void space in the porous medium. Apart from the relatively large Voronoi cells randomly distributed along the model, the distributions are essentially constant for the most part of the system, with an increase for the sub windows close to the top side of the system. However, experiments with relatively large generalised Bond numbers have a constant probability density of wetted cells along the top-bottom direction, which is reflected in the distribution points in Fig. 4(a) and 4(b). The correlation observed between probability density differences arises from the fact that generalised Bond number defines the stability of displacement structure. Additionally, as expected, the probability density of wetted cells tends to 0 for sub windows close to the outlet, since the corresponding region filled with liquid phase and saturated clusters are only formed in the area mostly occupied by the air phase.

Similarly, in Figs. 4(c) and 4(d) we divided the perpendicular distance into 36 equal sub windows (width of 5 mm), each spanning the height of the entire porous medium (230 mm out the whole micro-model width of 270 mm). The histograms show the probability density of wetted cells within sub windows. To quantify the variance of our data sets, the standard deviation (std) is shown in the plot legends in Figs. 4(c) and 4(d). The low standard deviation for each data set indicates that the distribution is essentially stable, which shows that the boundaries on the $x$-axis of the model have little effect on the formation of saturated clusters. In addition, as can be seen, the mean values of the wetted cell probability density decrease as the generalised Bond number increases.

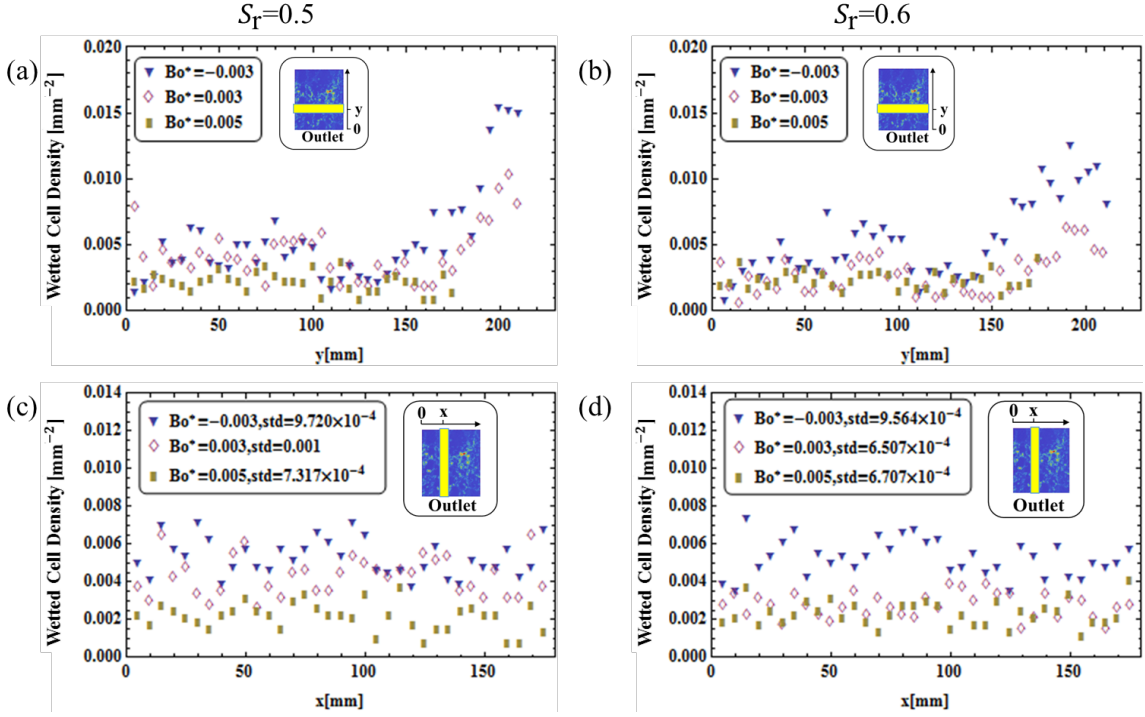

**Fig. 4** Histogram showing the spatial distributions of wetted cells in the model for three experiments in a Voronoi mesh: (a) and (b) the wetted cell density inside sub windows in the top-bottom direction for different saturation levels, as a function of the distance $y$ to the outlet; (c) and (d) the wetted cell density in the perpendicular direction for different saturation levels. The insets show the corresponding sub window orientations with respect to the model (made ≈ 4 times taller than the ones used in our analysis, to aid visualisation.

As can be seen in Fig. 4, the histograms in the first column ($S_r = 0.5$) have higher probability densities than those in the second column ($S_r = 0.6$) for the same sub windows in both directions, i.e., the lower saturation level has more wetted cells than the high saturation level at the same region in the model. This is due to the removal of the large fully saturated region behind the liquid front, and for smaller saturation levels larger accumulated saturated areas can be extracted. This can be explained by considering that saturated clusters form when the liquid front meets different conductive pathways during drainage processes, and therefore a sufficient distance along the flow direction is a prerequisite for the formation of such clusters. At lower saturation levels, a greater air content is present, more resistive pathways are bypassed and subsequently the accumulated area of viscous-detained liquid also increases. In addition, as $Bo^*$ increases, the accumulated wetted cell area included in saturated clusters is reduced. This follows the aforementioned theory about the influence of $Bo^*$ on geometry of liquid front in Section 2: for large negative values of $Bo^*$, the capillary fingers are widely spread along the liquid front during drainage processes; for large positive values of $Bo^*$, the liquid front is observed as a nearly flat geometry. For flat liquid fronts, gravity is always large enough to stabilise the displacement geometry leading to fewer saturated clusters trapped behind the liquid front and for unstable liquid fronts, more thin fingers can be observed, which are disconnected from the fully saturated region.

To look into the influence of packing characteristics on saturated cluster formation and evolution during drainage processes, we plot the number and probability distributions of wetted cells in Fig. 5. The wet cells are labelled with 1 and dry cells are labelled with 0. By collecting the amount of 1- labelled cells in an individual group divided by the total amount of the cells in a group, the wetted probability for each group is computed. The details can be found in Appendix A. In Fig. 5, we specified the bin space with 0.4 mm$^2$ and collected a list of bins and histogram heights of the values. The pink curves with empty circles in each image are computed from initially fully saturated images with different drainage conditions. The other coloured curves represent wetted cell area distributions within saturated clusters for different saturation levels associated with different drainage conditions. For these cases, the largest continuous saturated region below the liquid front is removed from every picture and only saturated clusters were left behind in each frame. By employing image processing operations on individual images, we collected the data stream from five cases with the same flow conditions and merged all sub-lists as the final step. To have a better visualisation of how the data is distributed, we depicted both the wetted cell area number distribution and wetted cell area probability distribution as functions of cell area.

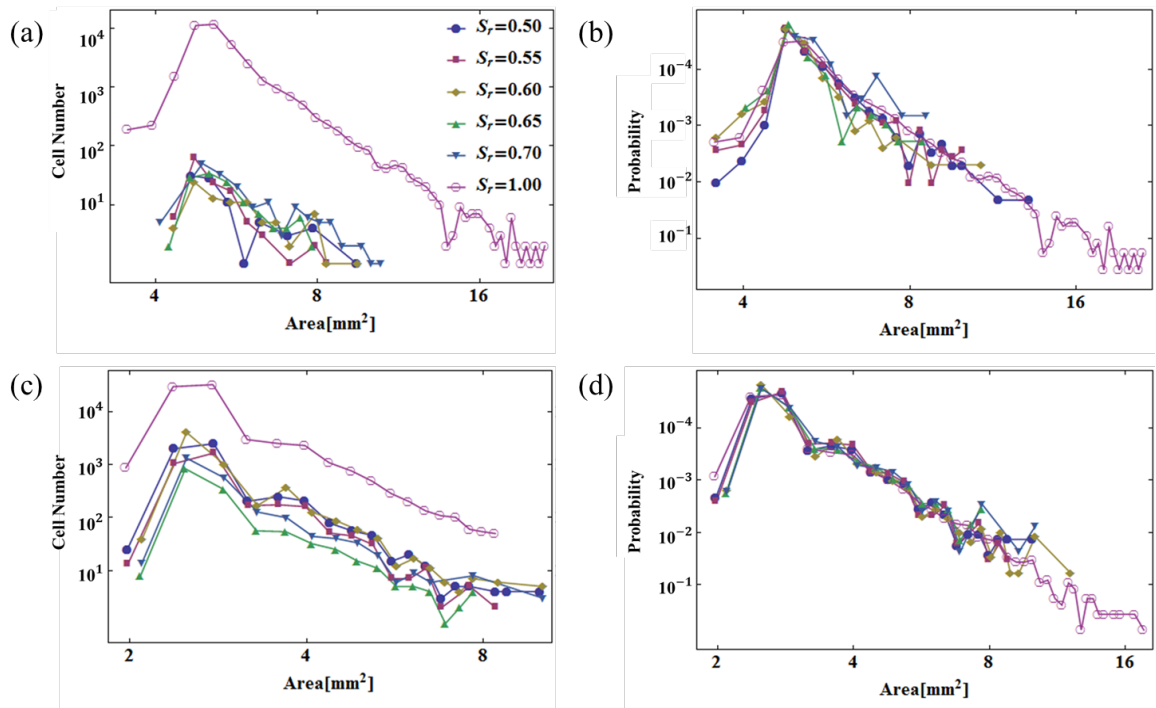

**Fig. 5** Wetted Voronoi cell and Delaunay Triangle (within saturated clusters) area distributions for six drainage conditions. The generalized Bond number in this case is −0.002, which could be referred to the experiments in Table. 1. The wetted cell number and probability distribution are calculated using Voronoi, (a) and (b), and Delaunay tesselations, (c) and (d).

In Figs. 5(a) and (b), it can be seen that over half of the cells have areas in the range 4-6 mm$^2$. By removing the projection area of individual glass beads ($\pi$ mm$^2$ in this study), the void space in these cells is then found to be 1.72-5.72 mm$^3$, which correlates to relatively close packed regions in the experimental sample. Table 2 shows the pore-scale size parameters and wetted probabilities for Voronoi cells computed from five pictures. The experiment shown here corresponds to a saturation level value of 0.5 and a generalized Bond number of −0.003. By grouping all cells with their individual face area, the corresponding dry/wet properties of every cell are also distributed into the same group. The relatively large values of wetted probability in Table 2 are most widely observed cells in the crystalized region. However, the smaller values are from cells with extremely large or small areas. This trend shows that in our porous medium, the most widely distributed cells have stronger liquid trapping ability than the extreme regions. Similarly, in Figs. 5(c) and (d), results have shown that more than half of the wetted cells are distributed in the relatively close packed region for all drainage conditions.

**Table 2** Pore-scale size parameters and wetted probabilities at different individual face area of Voronoi tessellation for $Bo^* = -0.003, S_r = 0.5$.

| | $Bo^* = -0.003$ | | |
|---|---|---|---|
| Area of Individual Cell (mm$^2$) | Wetted Probability (%) | Mean (mm$^2$) | Standard Deviation (mm$^2$) |
| 3.14< $A$ <4.14 | 0.57 | 3.69 | 0.30 |
| 4.14≤ $A$ <5.14 | 3.29 | 4.78 | 0.17 |
| 5.14≤ $A$ <6.14 | 2.32 | 5.51 | 0.26 |
| 6.14≤ $A$ <7.14 | 3.01 | 6.60 | 0.28 |
| 7.14≤ $A$ <8.14 | 1.33 | 7.57 | 0.28 |
| 8.14≤ $A$ <9.14 | 2.41 | 8.58 | 0.27 |
| 9.14≤ $A$ <10.14 | 4.33 | 9.59 | 0.25 |
| 10.14≤ $A$ <11.14 | 0.63 | 10.64 | 0.28 |

## 3.3 Saturated cluster distribution

To characterize the effect of drainage conditions on the evolution of saturated clusters, statistical analysis of saturated cluster geometries was conducted. The largest saturated cluster identified in each image, corresponds to the saturated region in front of the continuous liquid front, and is therefore excluded from the analysis. Here, we first look at the effective size of the saturated clusters, which is acquired by first collecting the number of pixels in a localized saturated region and then divided by the spatial resolution 30 pixels per pore (1 mm$^2$). To acquire sufficient data for statistical analysis, we performed five sets of experiments for each set of conditions. By grouping pictures with the same saturation, $S_r$, and generalized Bond number, $Bo^*$, and rearranging the experimental data into multiple sub-lists by a given bin size, the saturated cluster distributions are obtained.

The insert in Fig. 6(a) shows some observed elementary units of saturated clusters containing one to three glass beads during drainage processes. (I) In the subfigure (1) in Fig. 6(a), the central glass bead is fully coated and its neighbouring pores are occupied by the dyed liquid. In this subfigure, the saturated cluster is defined as the combination of the central glass bead and all its neighbouring pore spaces. (II) For saturated clusters with two glass beads inside, as shown in the subfigure (2) in Fig. 6(a), the volume of liquid in the inter-particle pore depends on the distance between two neighbouring beads, which can be quantified by the projection area of saturated clusters. (III) As for saturated clusters containing three beads, we assume that the pore structures in subfigure (3) in Fig. 6(a) are the same. However, owing to the fact that the capillary effect in granular media can be influenced by multiple factors, including pore size, flow rate and the gravitational force, the projection area of these two saturated clusters are different. These factors can be quantified by the generalized Bond number $Bo^*$. In what follows, saturated clusters are extracted for different values of $S_r$ and $Bo^*$. By collecting all the data at each stage and plotting their probability density distribution, it is possible to determine appropriate distribution parameters that represent the data set, and relate these to $Bo^*$.

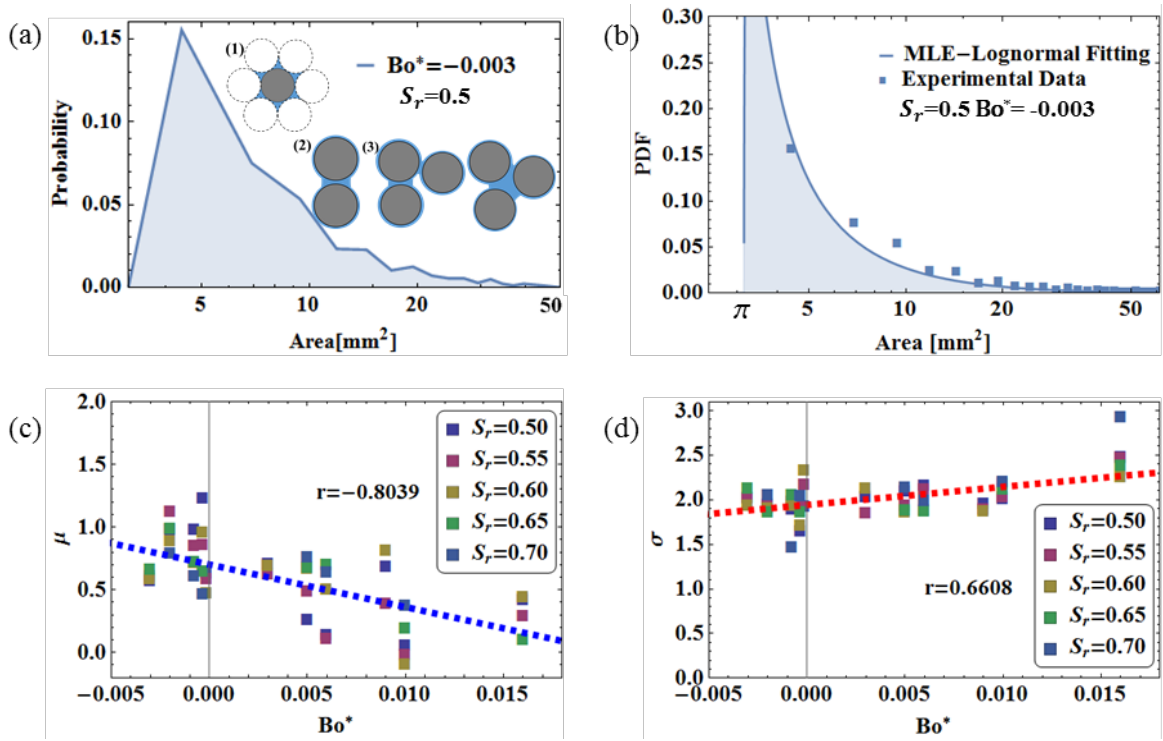

**Fig. 6** (a) The size distribution of saturated clusters for different liquid saturation levels with $Bo^*$ =-0.003. (b) Saturated cluster distribution at the same degree of saturation ($S_r$=0.5) with $Bo^*$ =-0.003 (Dots: the experimental data, and Line: lognormal distribution by maximum likelihood estimation). (c) Scale parameter $\mu$ and (d) shape parameter $\sigma$ as a function of generalised Bond number, $Bo^*$, for different saturation levels.

Fig. 6(b) is a typical probability distribution of saturated cluster areas with liquid saturation level $S_r = 0.5$. After grouping the dataset with unit width bins equal to 2.5 mm$^2$, the dataset is separated into sub lists, based on which the probability density distribution is calculated. It is observed that the probability density tends to the peak value (0.15-0.2 / mm$^2$) around 4-8 mm$^2$. As our porous medium is constructed by glass beads of 2 mm in diameter, the most frequent saturated clusters in the experimental cell can be interpreted as containing 1-2 glass beads. Different types of distributions have also been tested, while a lognormal distribution function, Eq. (1), was selected here to best represent the experimental data. The goodness-of-fit for three common fitting models are further compared with details provided in Appendix B.

As stated in Section 2.3, the lognormal density distribution function has two parameters, i.e., scale parameter $\mu$ and shape parameter $\sigma$, which govern the shape of the distribution curve. To estimate these two parameters from the given observations, we use the maximum likelihood estimation (MLE) method, described in Section 2.3. The resulting log-normal distribution can then be plotted by substituting the specific parameters acquired from the MLE method, as in Fig. 6(b). This figure illustrates the saturated cluster probability distribution as a function of area for different values of the generalized Bond number $Bo^*$ at a specific saturation levels $S_r$, and good agreement can be found between the MLE-fitted distribution and experimental data. Due to the removal of saturated regions that are smaller than a projection area of a typical glass bead ($\pi$ mm² in this study), the corresponding fitting curve in Fig. 6(b) starts at the point: ($\pi$, 0).

After collecting parameters $\mu$ and $\sigma$ for all series in the experiments, we applied Pearson's correlation coefficient method to quantify the relative linear relationship between them. To compute the correlation coefficient for both $\sigma$ and $\mu$ as a function of $Bo^*$, we substitute the acquired values of these two parameters with respect to $Bo^*$ into $r_{X,Y} = \frac{cov(X,Y)}{\sigma_X \sigma_Y}$. The correlated values for both parameters are shown in Figs. 6(c) and 6(d). In Fig. 6(c), the correlation coefficient of $\mu$ and $Bo^*$ is $-0.8039$, implying that all data points lay close to a line for which $\mu$ decreases as $Bo^*$ increases. On the contrary, the value of correlation coefficient of $\sigma$ and $Bo^*$ is $0.6608$, indicating that $\sigma$ increases as $Bo^*$ increases, as shown in Fig. 6(d).

For a given lognormal distribution, $\sigma$ and $\mu$ are the mean and standard deviations of the variable's natural logarithm, respectively. The linear relations obtained for the lognormal density function parameters as functions of $Bo^*$ are therefore correlated to the logarithmic mean and variation for saturated clusters sizes. In this case, as $Bo^*$ increases, the logarithmic mean saturated cluster size decreases; meanwhile, the saturated cluster sizes are spread out over a wider range of values. This correlation can be used to predict the geometrical distributions of saturated clusters trapped behind a liquid front for known pore structures and drainage conditions.

## 4. Conclusion

In this study we experimentally evaluated the formation of a temporarily entrapped liquid phase and its subsequent evolution during drainage processes of wetted granular materials. In our experiments the competition among viscous, gravitational and capillary forces was controlled to achieve different conditions of drainage from the porous media (characterised by the generalised Bond number, Bo*). Due to the inhomogeneous capillary threshold (governed by pore throat) in the porous medium and dominance of viscous effects, saturated clusters were trapped behind the moving front. The spatial and temporal distribution of saturated clusters can thus be observed, and the following conclusions can be made:

(1) The fraction of viscous limited pore bodies depends on the hydraulic properties of the resistive liquid phase, which is held in poorly connected pore clusters and small pore bodies due to the fact that gravitational force cannot overcome the effect of the viscous force. It is demonstrated that, under various drainage conditions and saturation stages, the statistical distribution of the liquid phase is determined by the intrinsic pore network characteristics. It is found that the pore-scale structure has a strong influence on the spatial extent of viscous delayed pores during a multiphase flow. The wetted cell size probability curve closely follows the global-scale cell size probability curve.

(2) The size of saturated clusters is found to follow a lognormal distribution, and the scale parameter $\mu$ and shape parameter $\sigma$ negatively and positively correlate with Bo*, respectively. Due to the fact that a relatively strong linear dependence exists between the lognormal distribution parameters, $\mu$ and $\sigma$, and generalised Bond number, Bo*, we are able to describe the cluster distribution at any given drainage conditions and saturation levels.

By extending the model towards three dimensional systems based on the explicit description of disordered pore networks, conclusions extracted from this study can be applied to a variety of situations in realistic geological scenarios involving multiphase flow. The ability to predict the presence of residual fluid phase left behind a moving liquid front on the basis of topological properties is of value in optimising water injection processes towards improved oil recovery (Blunt et al., 2002; Vishnudas and Chaudhuri, 2017) and pollutant removal (Saba and Illangasekare, 2000). The observations of the present study further show how particle scale affects overall flow conditions, potentially facilitating improved formulation of the effective stress term in unsaturated soil mechanics (Gan et al., 2013; Likos, 2014; Xu and Xie, 2011).

The variation of the capillary threshold here was limited, due to the topological homogeneity inherent to packings of monosized glass beads. For more complex systems, a quantitative prediction of the viscous limited saturated cluster distribution remains challenging. The results presented here provide an initial step towards a better understanding of drainage in porous media, suggesting further detailed modelling and experimental studies into saturated clusters formation and evolution.

# Acknowledgements

The authors are grateful for the financial support from Australian Research Council (Projects DE130101639 and DP170102886) and The University of Sydney SOAR Fellowship.

# Appendix A: Voronoi and Delaunay tessellation analysis

The procedures of using Voronoi and Delaunay tessellation in our statistical analysis are shown in detail below:

1. We first compute Delaunay triangles and Voronoi cells associated with the network using the image processing method mentioned in Section 3. After numbering and collecting the vertex coordinates of every cell in both mesh regions, we are then able to create region-member functions associated to each cell, for instance, the $i$-th region-member function for the $i$-th cell.
2. Then, import all pictures representing different saturation levels and drainage conditions into our calculation. These pictures are then be performed with morphological image processes and transformed into a black and white image, while the black and white pixels stand for dry and wetted regions, respectively. The black islands inside the white area are the wetted clusters trapped in the air during drainage. To allocate our focus on the saturated clusters behind the liquid front, we removed the largest continuous black island in each black and white picture, leaving other relatively small wetted clusters inside the pictures.
3. With the help of the built-in function (PixelValuePositions) in Mathematica, we then extracted all the black pixel positions in every picture. By repeatedly applying different black points into the $i$th region-member function, we returned to the results of true or false for whether these points are region members of the $i$th cell. This step gave us the results of the amount of points in the $i$th cell and we were then able to know the amount of black points in every cell by repeatedly employing this step on all rest cells in an individual picture. As a result, we were able to return to a list of number of black points in each cell aligned with the cell number.
4. By comparing the ratio of black pixel number to white pixel number, we classified the cells into two groups: dry and wet. After filtering the uncorrelated information from the aforementioned steps, we finally came to a list giving the area and wet (1) / dry (0) properties for every cell in an individual picture.

To enlarge our dataset, we also did five experiments under same conditions and merged all five lists to perform further statistical analysis.

# Appendix B: Goodness-of-fit test results

Table. B1 shows two widely used tests, the Pearson's $\chi^2$ and *Anderson-Darling* methods to test the validity of three assumed distribution models (Lognormal distribution, Weibull distribution and Gamma distribution). We applied a particular dataset ($S_r$ =0.5; $Bo^* = -0.003$) to the two test methods and listed the resulting values in Table. B1. These results are used to verify the validity of lognormal distribution which has been selected on the basis of other prior considerations.

**Table. B1** Obtained maximum likelihood parameters from a specific data-set. Chi-square test and *Anderson − Darling* criterion are used to quantify the goodness of fit.

| $S_r$=0.5 $Bo^*$=-0.003 | Parameter | | Pearson's $\chi^2$ | | *Anderson-Darling* | |
|---|---|---|---|---|---|---|
| | | | Statistic | P-value | Statistic | P-value |
| Lognormal Distribution | $\mu = 1.078$ | $\sigma = 1.590$ | 23.841 | 0.690 | 0.582 | 0.665 |
| Weibull Distribution | $\alpha = 0.603$ | $\beta = 6.538$ | 97.652 | $1.213 \times 10^{-9}$ | 11.755 | $3.019 \times 10^{-4}$ |
| Gamma Distribution | $\alpha = 0.463$ | $\beta = 25.444$ | 202.537 | $2.267 \times 10^{-9}$ | 32.2891 | 0.000 |